\documentclass[aps,prl,reprint,showpacs,showkeys,superscriptaddress,preprintnumbers,groupedaddress]{revtex4-1}
\usepackage{amsmath,amssymb,bm,mathrsfs}
\usepackage{srcltx}
\usepackage{dsfont}
\usepackage{epsfig}
\usepackage{slashed}
\usepackage{bbold}
\usepackage{psfrag}
\usepackage{xcolor}
\PassOptionsToPackage{caption=false}{subfig}
\usepackage{subfig}
\usepackage{xfrac}
\usepackage{multirow}
\usepackage{booktabs}
\usepackage[colorlinks=true,linkcolor=blue,citecolor=blue,urlcolor=violet]{hyperref}

\newcommand{\be}{\begin{eqnarray}}
\newcommand{\ee}{\end{eqnarray}}

\newcommand{\eq}{{\textrm{eq}}}

\newcommand{\nn}{\nonumber}
\def\({\left(}
\def\){\right)}

\definecolor{colorRTD}{rgb}{.2,.2,.7}

\usepackage{soul}

\begin{document}

\title{Thermal Relic Targets with Exponentially Small Couplings} 

\author{Raffaele Tito D'Agnolo}
\email{dagnolo@slac.stanford.edu}
\affiliation{SLAC National Accelerator Laboratory, 2575 Sand Hill Road, Menlo Park, CA, 94025, USA.} 

\author{Duccio Pappadopulo}
\email{duccio.pappadopulo@gmail.com}
\affiliation{Bloomberg LP, New York, NY 10022, USA.
} 

\author{Joshua T. Ruderman}
\email{ruderman@nyu.edu}
\affiliation{
Center for Cosmology and Particle Physics,
Department of Physics, New York University, New York, NY 10003, USA.
} 

\author{Po-Jen Wang}
\email{pjw319@nyu.edu}
\affiliation{
Center for Cosmology and Particle Physics,
Department of Physics, New York University, New York, NY 10003, USA.
} 

\begin{abstract}
If dark matter was produced in the early Universe by the decoupling of its annihilations into known particles, there is a sharp experimental target for the size of its coupling.  We show that if dark matter was produced by inelastic scattering against a lighter particle from the thermal bath, then its coupling can be exponentially smaller than the coupling required for its production from annihilations.  As an application, we demonstrate that dark matter produced by inelastic scattering against electrons provides new thermal relic targets for direct detection and fixed target experiments.
\end{abstract}

\maketitle


{\bf Introduction---} Dark Matter (DM) was produced in the early Universe by unknown dynamics.  If the production of DM is tied to its measurable interactions, experiments can help disentangle its cosmological origin.  The most studied example is the Weakly Interacting Massive Particle (WIMP)~\cite{Lee:1977ua, Kolb:1990vq, Gondolo:1990dk, Jungman:1995df}, with abundance set by the decoupling of its annihilations into Standard Model (SM) particles.
In this case the DM annihilation rate is predicted and testable by current experiments.

\begin{figure}[!!!t]
\begin{center}
\includegraphics[width=0.5 \textwidth]{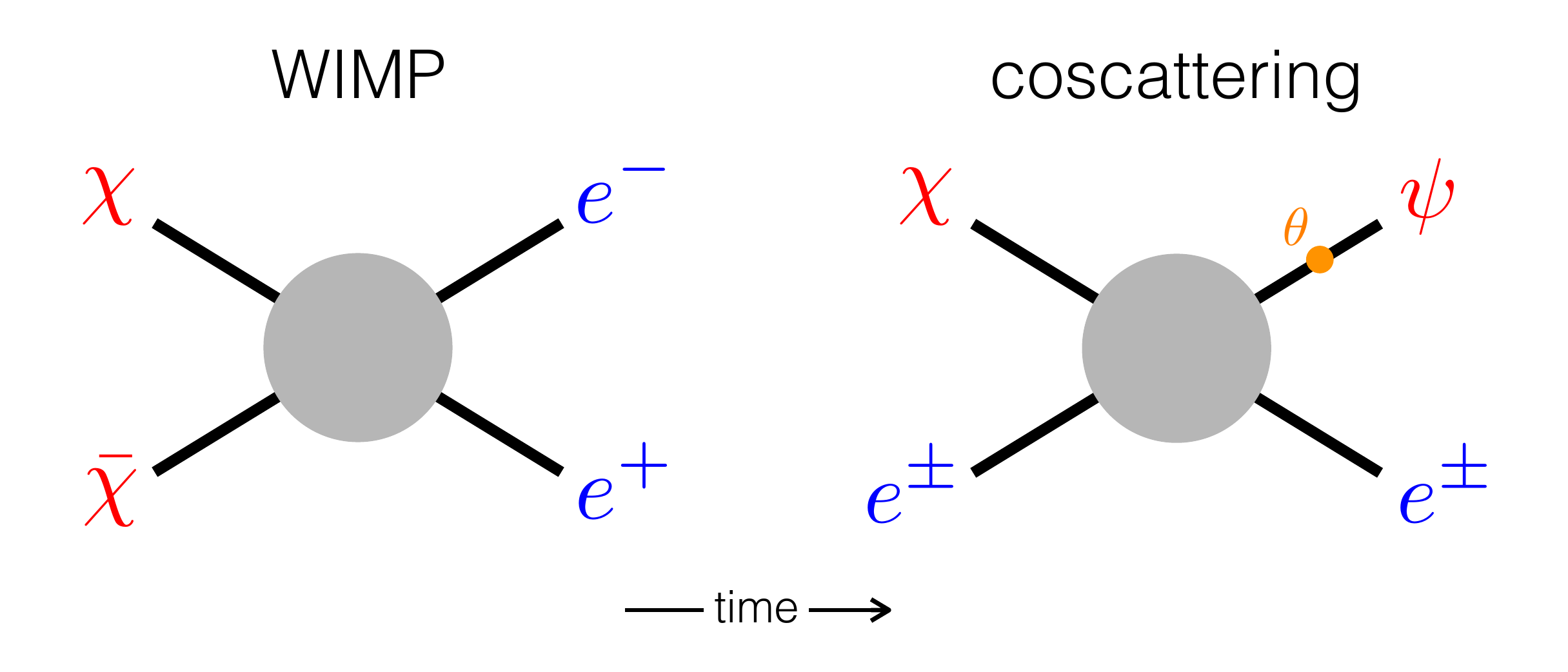}
\end{center}
\vspace{-.3cm}
\caption{\small 
The process setting the relic density for the WIMP (left) and coscattering (right), where $\chi$ is DM and $\psi$ is a dark partner. \label{fig:scheme}}
\end{figure}

A well-known variant of the WIMP is DM with a mass in the MeV to GeV range that annihilates into electrons and positrons (left of Fig.~\ref{fig:scheme})~\cite{Boehm:2003hm,Pospelov:2007mp}. This scenario requires additional force mediators beyond the SM~\cite{Lee:1977ua,Boehm:2003hm,Pospelov:2007mp,Strassler:2006im,ArkaniHamed:2008qn}\@. The requirement that DM has the observed abundance fixes the size of its coupling to electrons, implying a sharp target for direct detection and fixed target experiments~\cite{Izaguirre:2014bca,Izaguirre:2015yja,Essig:2015cda,Alexander:2016aln,Battaglieri:2017aum,Berlin:2018bsc}. 
The WIMP and its variants  are now driving a large experimental effort.
Are there alternate scenarios on the same theoretical footing as the WIMP that experiments should also target?

In this letter we consider {\it thermal relic targets} that share the attractive features of the WIMP\@.  We seek cosmologies where DM (1) begins in thermal equilibrium with a temperature that tracks the photon temperature, (2) follows a standard cosmological history, and (3) has abundance determined by the decoupling of an interaction between DM and SM particles.  The first two assumptions are violated by non-thermal DM~\cite{Moroi:1993mb,Dodelson:1993je, Feng:2003xh, Hall:2009bx}, cannibalism~\cite{Carlson:1992fn,Kuflik:2015isi,Pappadopulo:2016pkp,Farina:2016llk}, and late entropy production~\cite{Gelmini:2006pw, Wainwright:2009mq, Berlin:2016vnh}; the third is not satisfied if DM annihilates into hidden sector states~\cite{Finkbeiner:2007kk,Pospelov:2007mp,Feng:2008ya,Hochberg:2014dra,Hochberg:2014kqa,Evans:2017kti} or has abundance set by a primordial asymmetry~\cite{Kaplan:2009ag,Petraki:2013wwa,Zurek:2013wia}.

\begin{figure*}[!t]
\begin{center}
\includegraphics[width=0.9 \textwidth]{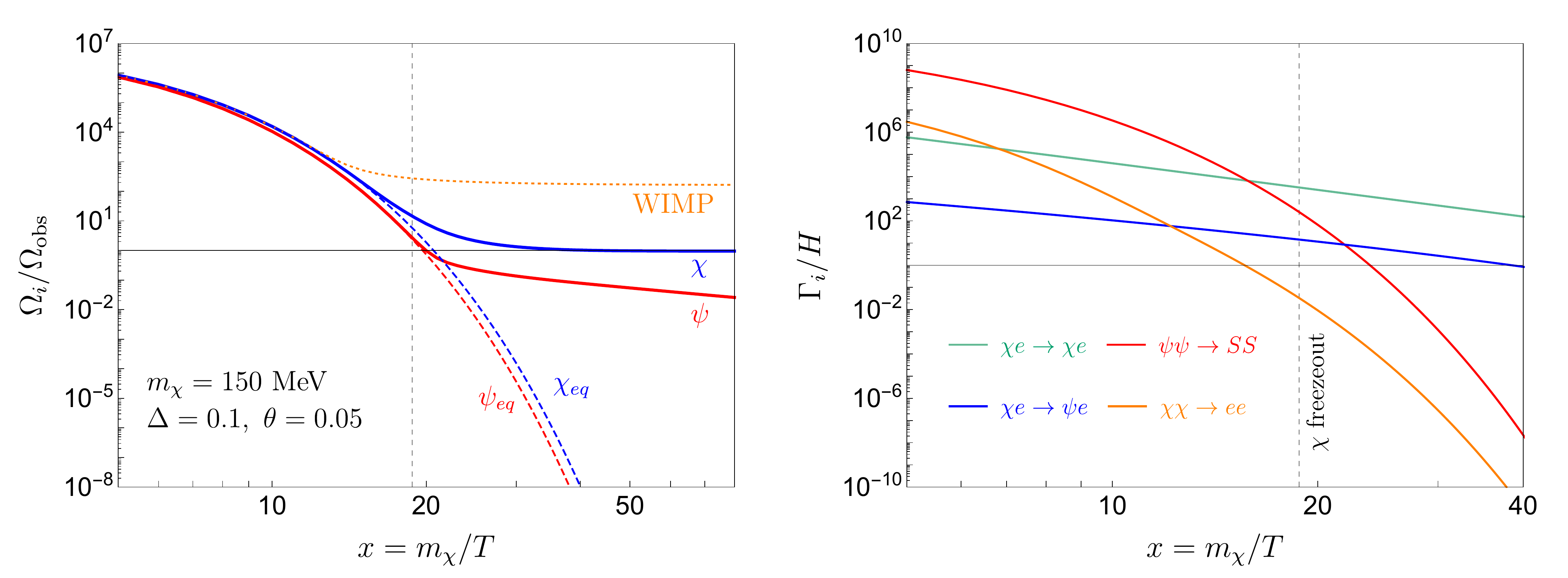}
\end{center}
\vspace{-.3cm}
\caption{\small {\it Left panel}: Evolution of the DM energy density for an ordinary WIMP (orange dotted) and coscattering (solid blue). The red curve describes freeze-out of the DM partner $\psi$. {\it Right panel}: thermal rates of the interactions relevant for coscattering as a function of $x = m_\chi / T$.  We plot the momentum transfer rate for elastic scattering (green) and $(n_\psi^{\rm eq})^2 \left< \sigma v \right>/n_\chi^{\rm eq}$ for $\psi$ annihilations (red).  In both plots, $m_\chi=150\text{ MeV},~\Delta=0.1,~\theta=0.05, ~\epsilon=3\times10^{-5},~\lambda_{\psi S}=0.0064$, $\alpha_d=0.5$, $m_S=156\text{ MeV}$, and $m_A=3m_\chi$. 
\label{fig:Omega}}
\end{figure*}

We identify a new example that satisfies the above assumptions: DM, $\chi$, is produced by scattering against electrons.  Elastic scattering, $\chi e^\pm \rightarrow \chi e^\pm$, does not change the number of DM particles, but the abundance can be set by inelastic scattering, $\chi e^\pm \rightarrow \psi e^\pm$  (right of Fig.~\ref{fig:scheme}), where $\psi$ is a dark partner that experiences its own rapid annihilations.  This is an example of {\it coscattering}~\cite{DAgnolo:2017dbv} as  studied recently by Refs.~\cite{Garny:2017rxs,Cheng:2018vaj,Garny:2018icg, Garny:2019kua, Kim:2019udq}.  See also Refs.~\cite{Farrar:1995pz,Chung:1997rq,Pospelov:2010cw} for related prior work. 
Here we introduce a mechanism for extending any WIMP-like model to include a coscattering phase, and we show how this opens up new parameter space with smaller DM coupling.

The rate for DM to coscatter (annihilate) is proportional to the electron (DM) number density $n_e$ ($n_\chi$) multiplied by the relevant cross section.  Suppose that $m_e < T_f < m_\chi$, where $T_f$ is the temperature at which annihilations or coscatterings decouple.  Then the number density of electrons, which are relativistic, is larger than the number density of DM, which is non-relativistic, by the exponentially large factor $n_e / n_\chi \propto e^{m_\chi/T_f}$.  Therefore, the rate of coscattering can exceed annihilation despite an exponentially smaller cross section, opening up vast new parameter space for thermal relic targets.  Below we focus on a particular model with coscattering against electrons, but our observation applies to many more models and to replacing the electron with other SM particles.  Another scenario that satisfies the above assumptions and leads to a smaller DM coupling is annihilation through a pole~\cite{Feng:2017drg}, requiring a special relation between the mass of DM and another particle.

The rest of this letter is organized as follows.  We begin by comparing coscattering to the more familiar WIMP, and we derive the exponential factor that enlarges the parameter space for coscattering.  We then show how any model with annihilating DM can be extended to include coscattering, taking as an example scalar DM that couples to electrons through a dark photon.  We explore the detailed phenomenology of this example, highlighting the experimentally testable parameter space.  Our Appendix contains a map of the phase space of this model, charting where the the DM abundance is set by coscattering or alternate mechanisms.

{\bf Coscattering vs. Annihilations--- } 
A WIMP with mass $m_\chi$ begins in equilibrium with the SM thermal bath when $T> m_\chi$. Its abundance is depleted through annihilations when $T< m_\chi$. 
When the rate of DM annihilations becomes slower than the expansion of the Universe
\be
n_\chi^{\rm eq} \langle \sigma v\rangle \approx H\, ,  \label{eq:suddenWIMP}
\ee
the total number of DM particles is fixed (freezes-out) determining the observed relic density today~\cite{Kolb:1990vq, Gondolo:1990dk}. Here we have used the superscript ``${\rm eq}$" to denote particles in thermal equilibrium with the SM bath. The sudden freeze-out approximation described above points to annihilation cross sections comparable to those induced by the SM weak interactions:
\be
\frac{\Omega_{\rm WIMP}}{\Omega_{\rm DM}}\approx \frac{0.3\;{\rm pb}}{\langle \sigma v \rangle} \frac{x_f\sqrt{g_*}}{g_{*S}}\equiv \frac{\sigma_{\rm WIMP}}{\langle \sigma v \rangle}\, , \label{eq:relicWIMP}
\ee
where $x_f = T_f / m_\chi$ describes the temperature when Eq.~\ref{eq:suddenWIMP} is satisfied.

If annihilations are too small, and no other process removes DM, the relic density is too large.
In this regime, coscattering is an alternative mechanism for setting the DM abundance~\cite{DAgnolo:2017dbv}.  DM still begins in equilibrium with the SM thermal bath when $T> m_\chi$. The relic density is set when $T< m_\chi$ and inelastic scattering becomes slower than Hubble.  In the following we assume that DM scatters against an electron or positron, as on the right of Fig.~\ref{fig:scheme}, but other choices are possible~\cite{DAgnolo:2017dbv}.  DM scatters into a dark partner $\psi$ with a heavier mass, $m_\psi\geq m_\chi$, whose annihilations leave equilibrium after the inelastic scattering.  DM is depleted by $\chi e^\pm \rightarrow \psi e^\pm$ followed by $\psi$ annihilations.  When $\chi e^\pm \rightarrow \psi e^\pm$ freezes-out, so does the $\chi$ abundance.    The small component of $\psi$ left over from the later freeze-out of $\psi$ annihilations is converted into $\chi$ by decays, or else persists until today.

The main qualitative difference between coscattering and the WIMP  is that we can obtain the correct relic density with $\chi$ interactions that are exponentially smaller than required for annihilations.
If we assume that momentum transfer via $\chi e^\pm \to \chi e^\pm$ is efficient until after the coscattering diagram freezes-out, we can write a Boltzmann equation for the total $\chi$ number density
\be \label{eq:BE}
\frac{d n_\chi}{dt}+3 H n_\chi = -n_{e}^{\rm eq} \langle \sigma_{\chi \to \psi} v\rangle\left[n_\chi - n_\chi^{\rm eq}\right] \, .
\ee
In this case freeze-out occurs when
\be
n_{e}^{\rm eq} \langle \sigma_{\chi \to \psi} v\rangle \approx p H\, ,  \label{eq:sudden}
\ee
where $p \approx 10$ is chosen to match numerical solutions to Eq.~\ref{eq:BE}.
If we use detailed balance to write the thermal average for the endothermic process $\chi \to \psi$ in terms of the thermal average for the inverse process~\cite{DAgnolo:2017dbv}, 
we can write the $\chi$ relic density as
\be\label{omegaratio}
\frac{\Omega_\chi}{\Omega_{\rm DM}}\approx 0.2 \frac{\sigma_{\rm WIMP}}{\langle \sigma_{\psi \to \chi} v \rangle}\,\frac{p\, g_\chi^2 \, x_f^{3/2}  e^{- x_f(1-\frac{\Delta m}{m_\chi}) }}{g_\psi (1+{\Delta m}/{m_\chi})^{3/2}}\, , \label{eq:relic}
\ee
where $\Delta m \equiv m_\psi - m_\chi$ and $g_{\chi,\psi}$ counts the number of degrees of freedom of $\chi$ and $\psi$.
In order to reproduce $\Omega_\chi=\Omega_{\rm DM}$ we require $\langle \sigma_{\psi \to \chi} v \rangle \sim e^{- x_f(1-\Delta m/m_\chi)} \sigma_{\rm WIMP}$.
This corresponds to a significant exponential suppression, as long as $\Delta m/m_\chi < 1$, because $x_f \sim 20$ is needed to match the observed abundance. (The necessary freeze-out temperature is universal across different types of thermal relics including coscattering and the WIMP)\@.


{\bf The Model---}  The coscattering phase can be added to any WIMP model in a modular way. Consider a WIMP with an annihilation channel, such as the left of Fig.~\ref{fig:scheme}, and add a mass mixing between DM and a dark partner, $\psi$, with its own rapid annihilations.  The coscattering diagram on the right of Fig.~\ref{fig:scheme} is generated by rotating the annihilation diagram and inserting the mixing.

As an example we consider dark scalar QED~\cite{Boehm:2003hm,Pospelov:2007mp,Izaguirre:2014bca,Izaguirre:2015yja,Essig:2015cda,Alexander:2016aln,Battaglieri:2017aum,Berlin:2018bsc} with DM coupled to the SM via a $U(1)_d$ massive dark photon, $A_d$, kinetically mixed~\cite{Holdom:1985ag} with the ordinary photon, $A$:
\be
\mathcal{L}&\supset&-\frac{1}{4}F_{\mu\nu}^2-\frac{1}{4}(F^{\mu\nu}_d)^2-\frac{\epsilon}{2}F^{\mu\nu}_dF_{\mu\nu}-\frac{m_A^2}{2} A_d^2 \, ,
\ee
where $F_{(d)}$ is the ordinary (dark) photon field strength, and $\epsilon$ is a dimensionless measure of the mixing.
DM is a complex scalar $\chi$ with charge 1 under $U(1)_d$: $\mathcal{L}\supset |D_\mu \chi|^2-\overline m_\chi^2|\chi|^2$.  DM can annihilate: $\chi \chi^* \rightarrow \gamma_d^* \rightarrow e^+ e^-$.

We add a complex scalar, $\psi$, that is neutral under $U(1)_d$ and mixed with $\chi$:
\be
\mathcal{L}\supset |\partial_\mu \psi|^2-\overline m_\psi^2|\psi|^2-\delta m^2\left( \chi^* \psi + {\rm h.c.}\right)\, . 
\ee
Coscattering, $\chi e \rightarrow \psi e$, is generated by the mixing. Quartic couplings for $\chi$ and $\psi$ can also be included without modifying our discussion.
The mass mixing $\delta m^2$ and the dark photon mass $m_A^2$ can arise from a dark Higgs coupled to $\chi$ and $\psi$.  
Note that we use $\overline m_{\chi, \psi}^2$ for mass parameters in the Lagrangian and $m_{\chi, \psi}^2$ for mass eigenvalues. The angle that rotates from the Lagrangian basis to the mass eigenstate basis is $\theta\sim \delta m^2/m_\chi^2$.  For the rest of this letter we (slightly) abuse notation by calling the lightest eigenstate $\chi$ and the heaviest $\psi$.

We take $\psi$ to annihilate to a dark state (this is a key difference versus Refs.~\cite{Garny:2017rxs,Garny:2018icg, Garny:2019kua}, where the dark partner has a large coupling to the SM), $\psi \psi^* \rightarrow SS$, where $S$ is a real scalar lighter than $\psi$,
\be
\mathcal{L}\supset \frac{1}{2}(\partial_\mu S)^2-\frac{m_S^2}{2}S^2-\frac{\lambda_{\psi S}}{2}S^2|\psi|^2-y_{eS}\left(S e^c e +{\rm h.c.}\right)\, .\, \nn 
\ee
$\chi$ couples to $S$ only through its mixing with $\psi$.
We take $S$ to decay rapidly to the thermal bath via a small Yukawa coupling to electrons.     There is a large range of values for this coupling ($10^{-4} \lesssim y_{eS} \lesssim 10^{-9}$) such that $S$ remains in equilibrium with the SM bath during freeze-out but otherwise $y_{eS}$ does not enter the relic density calculation.

We consider parameters where  $\chi\chi^* \to  e^+ e^-$ annihilations are too feeble to set the relic density.  However $\psi\psi^* \to SS$ annihilations are efficient and the last process to freeze-out. The coscattering process $\chi e \to \psi e$ leaves equilibrium after $\chi\chi^* \to  e^+ e^-$, but before $\psi\psi^* \to SS$.  It is the fastest process converting $\chi$ to $\psi$ and its freeze-out determines the $\chi$ abundance. This hierarchy of thermal rates is depicted in Fig.~\ref{fig:Omega}. The figure also shows that $\chi$ would have too large of an abundance in the WIMP limit where only $\chi$ annihilations are active, while the addition of $\psi$ leads to the correct relic density via coscattering.

The freeze-out hierarchy described in the previous paragraph can be realized with the spectrum: $m_A > m_\psi > m_S > m_\chi \gtrsim \sqrt {\delta m^2}$ and the hierarchy of couplings: $g_d =\mathcal{O}(1)\gg \epsilon, \lambda_{\psi S}$. Furthermore, we fix $r_S \equiv (m_S-m_\chi)/(m_\psi-m_\chi) = 0.75$ in the following. More details on these choices and the freeze-out phases of this theory can be found in the Appendix. 

\begin{figure}[!!!t]
\begin{center}
\includegraphics[width=0.45 \textwidth]{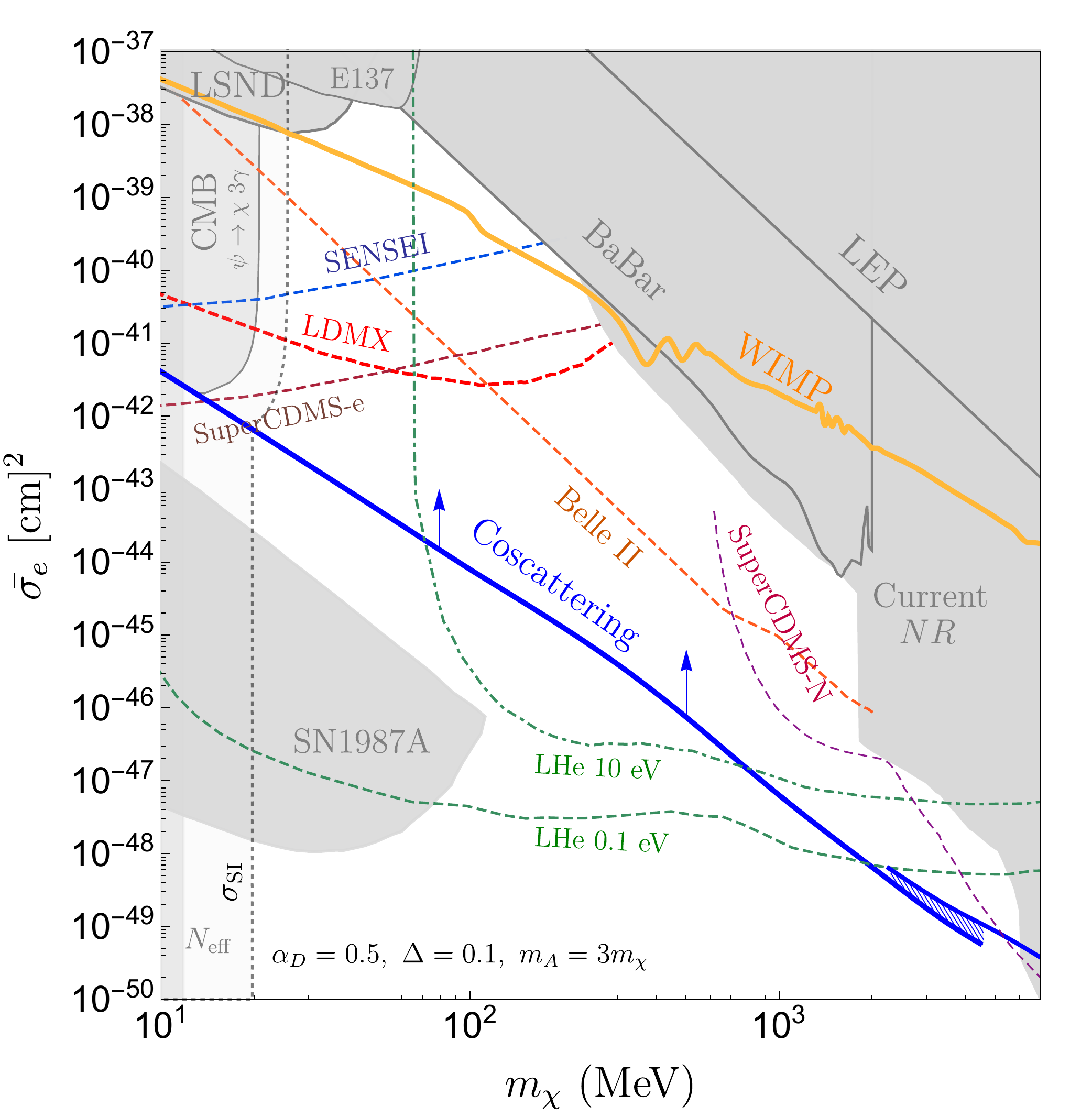}
\end{center}
\vspace{-.3cm}
\caption{\small Existing constraints and reach of future experiments as a function of the DM mass $m_\chi$, expressed in terms of the direct detection cross section for electron recoils $\bar \sigma_e$. In the region between the orange and the blue lines the observed relic density is obtained via the coscattering mechanism. In the plot we fix $\alpha_d=0.5$,  $m_A=3 m_\chi$, $r_S=0.75$, and $\Delta \equiv (m_\psi^2 - m_\chi^2)/m_\chi^2 = 0.1$.
\label{fig:pheno}}
\end{figure}

\begin{figure*}[!!!t]
\begin{center}
\includegraphics[width=0.9 \textwidth]{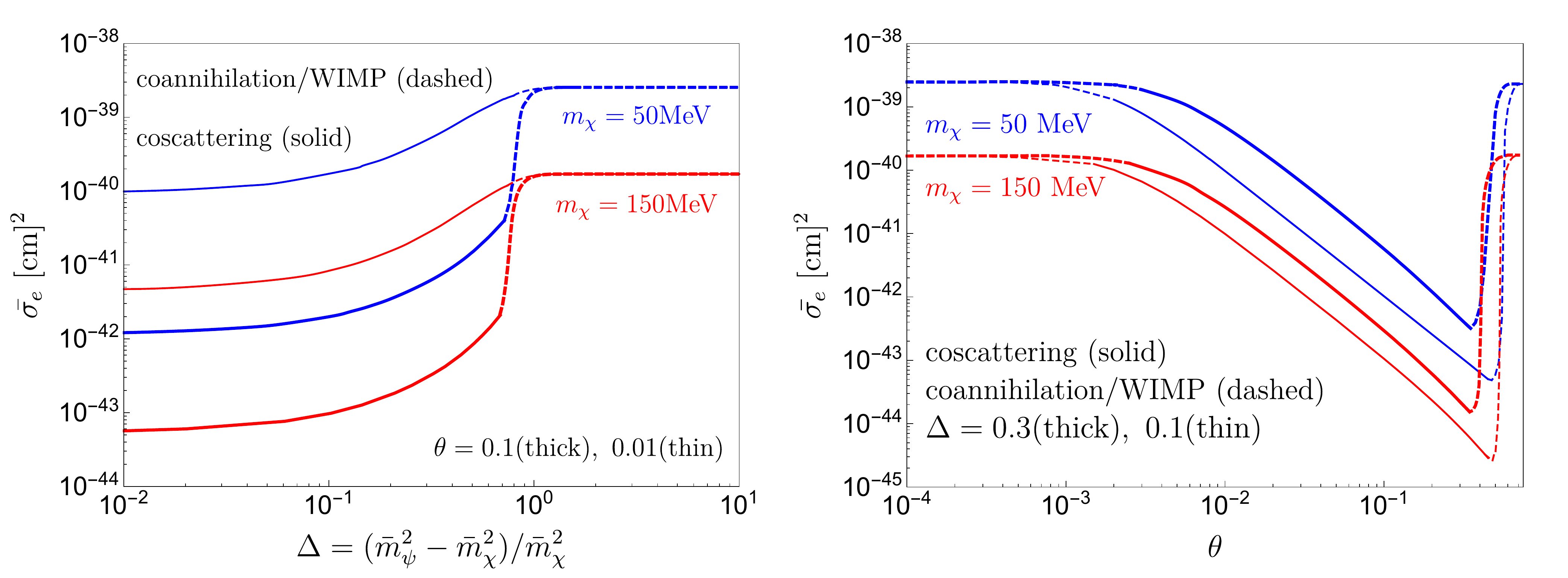}
\end{center}
\vspace{-.3cm}
\caption{\small Direct detection cross section for scattering off atomic electrons $\bar \sigma_e$ as a function of the mass splitting between DM and its partner $\psi$ (left panel) and their mixing (right panel), after the DM relic density is fixed to its observed value. In both plots we fix $\alpha_d=0.5$, $m_A=3 m_\chi$, and $r_S=0.75$.
\label{fig:sigma}}
\end{figure*}

{\bf Phenomenology---} The main qualitative feature of coscattering with SM states is the exponentially small coupling to SM particles, compared to the WIMP\@.  In Fig.~\ref{fig:pheno} we plot current constraints and future probes in terms of the scattering cross section relevant to electron recoil experiments: $\bar \sigma_e \approx e^2g_D^2\epsilon^2 m_e^2 c_\theta^4 /(\pi m_A^4)$. Coscattering can reproduce the observed relic density in the region between the solid orange line (where it reduces to WIMP freeze-out) and the solid blue line. In the coscattering region we find that $\chi$ is in kinetic equilibrium with the SM during freeze-out due to rapid energy exchange from $\chi e^\pm \rightarrow \chi e^\pm$~\cite{Bringmann:2006mu,Adshead:2016xxj}.
 This is distinct from previous studies of coscattering where DM kinetically decouples during freeze-out~\cite{DAgnolo:2017dbv,Garny:2017rxs,Cheng:2018vaj,Garny:2018icg,Garny:2019kua}.

Proposed experiments have the potential to probe a large fraction of the coscattering region, as shown in Fig.~\ref{fig:pheno}.  They include electron recoil direct detection experiments such as SENSEI~\cite{Tiffenberg:2017aac, Abramoff:2019dfb, Crisler:2018gci, Battaglieri:2017aum} and SuperCDMS-G2+~\cite{Battaglieri:2017aum}; nucleon recoil direct detection including SuperCDMS iZIP detectors~\cite{Battaglieri:2017aum} and superfluid helium detectors~\cite{Hertel:2018aal} (shown for 1 kg-yr exposure with a 10 eV energy threshold and 10 kg-yr with 0.1 eV); missing momentum fixed target experiments such as LDMX~\cite{Berlin:2018bsc,Akesson:2018vlm}; and high-luminosity electron-positron colliders such as Belle-II~\cite{Kou:2018nap,Berlin:2018bsc}. We also note that  NA64~\cite{Gninenko:2019qiv} and other proposed direct detection experiments~\cite{Battaglieri:2017aum} have potential sensitivity to our parameter space. 

In Fig.~\ref{fig:pheno} we fix $\alpha_d=0.5$, $m_A=3 m_\chi$, $r_S=0.75$, and $\Delta \equiv (m_\psi^2 - m_\chi^2)/m_\chi^2 = 0.1$. Within the coscattering region $\epsilon$ is determined by the value of $\bar \sigma_e$ on the $y$-axis, once $\theta$ is fixed at each point to give the right relic density. In this region the quartic $\lambda_{\psi S}$ is chosen at each point to select the coscattering regime, as discussed in the Appendix. Below the coscattering region the mixing angle is set to its maximum value $\theta=0.45$ that is reached on the blue line at the boundary of the coscattering parameter space.

The reason why coscattering provides a range of viable couplings as opposed to the WIMP relic density line can be understood with the help of Fig.~\ref{fig:sigma}. The relic density is set by processes that interchange $\chi$ and $\psi$, so is sensitive to the mixing angle, $\theta$, as shown in the right panel of Fig.~\ref{fig:sigma}. Larger $\theta$ allows for a smaller coupling of $\chi$ to electrons when fixing the relic density. So the smallest possible coupling of DM to the SM is achieved when $\theta=\mathcal{O}(1)$. Smaller values of $\theta$ span the region between the blue and the orange lines of Fig.~\ref{fig:pheno}. When $\theta$ becomes too large we enter a coannihilation or WIMP regime, depending on the parameters in the dark sector, as shown in Fig.~\ref{fig:sigma} and discussed in the Appendix.

We also comment on the $\chi-\psi$ mass splitting, $\Delta$. The coscattering process is endothermic ($m_\psi\geq m_\chi$) and the relic density depends exponentially on $\Delta$. This is shown in the left panel of Fig.~\ref{fig:sigma} and in Eq.~(\ref{eq:relic}). When $\Delta$ goes to zero there is no suppression from the thermal average and $\bar \sigma_e$ can be small, while for larger $\Delta$ a larger coupling of $\chi$ to electrons has to compensate the kinematical suppression. So $\Delta \approx 0$ gives the smallest direct detection cross section compatible with the observed relic density.
In Fig.~\ref{fig:pheno} we have chosen $\Delta = 0.1$; the smallest possible $\bar \sigma_e$ obtained for $\Delta=0$ is approximately a factor of 5 below the solid blue line in the plot.

We would  like to draw attention to the relative smoothness  of the line bounding the coscattering parameter space from below in Fig.~\ref{fig:pheno}, compared to the WIMP relic density line. The WIMP line is determined by $s$-channel DM annihilations and reflects the structure of SM resonances that occur when $2 m_\chi \approx m_{\rm SM}$.  Coscattering is $t$-channel and receives contributions from all SM states that are relativistic at freezeout, $T_f \gtrsim m_{\rm SM}$.  As the DM mass increases, new SM states contribute smoothly.
There is theoretical uncertainty when freeze-out happens near the QCD phase transition, $T_f \sim 100-200$~MeV\@.  In this regime we show both coscattering off quarks (upper curve) and pions/kaons (lower curve).

In Fig.~\ref{fig:pheno} we also show current constraints on our parameter space (shaded in gray). We display CMB bounds on $\psi \rightarrow \chi + 3\gamma $ decays~\cite{Slatyer:2016qyl} and $N_\text{eff}$~\cite{Aghanim:2018eyx}; supernova cooling~\cite{Chang:2018rso}; direct detection from nuclear recoils~\cite{Abdelhameed:2019hmk,Agnese:2015nto,Agnes:2018ves,Cui:2017nnn,Aprile:2018dbl}; BaBar missing energy searches~\cite{Lees:2017lec}; LEP measurements at the Z-pole~\cite{Hook:2010tw}; beam dump experiments such as LSND~\cite{Auerbach:2001wg,deNiverville:2011it} and E137~\cite{Bjorken:1988as, Batell:2014mga}; and DM self-interactions (we require $\sigma_{\rm SI}/m_\chi\gtrsim 1\,{\rm cm^2}/{\rm g}$)~\cite{Rocha:2012jg, Peter:2012jh, Markevitch:2003at, Clowe:2003tk, Randall:2007ph, Harvey:2015hha}. 


{\bf Conclusions---} In this paper we introduced a thermal relic that shares the attractive theoretical properties of the WIMP, but reproduces the DM relic density with an exponentially smaller coupling to the SM\@. We have shown that coscattering~\cite{DAgnolo:2017dbv} can be realized by extending any model of WIMP DM, opening up orders of magnitude of new parameter space.  We have explored the phenomenology of one concrete example model, providing a new benchmark for future light DM experiments. Our work provides further motivation for direct detection experiments that probe sub-GeV DM scattering with nucleons or electrons, and missing momentum searches at fixed target experiments.  Coscattering motivates extending the sensitivity of these programs beyond traditional WIMP targets.

\vspace{.3cm}
\begin{acknowledgements}
{\em Acknowledgements---} 
We would like to thank A. Berlin, P. Schuster, and N. Toro  for useful discussions. RTD is supported by DOE grant DE-AC02-76SF00515. JTR is supported by NSF CAREER grant PHY-1554858.
\end{acknowledgements}
\appendix
\section{Appendix}

{ \bf Analytical Estimates: the Coscattering Regime---} In this section we provide an analytical understanding of the model introduced in the paper, discussing the parametric relations defining the boundaries of its various freeze-out phases.

\begin{figure}[!!!t]
\begin{center}
\includegraphics[width=0.45 \textwidth]{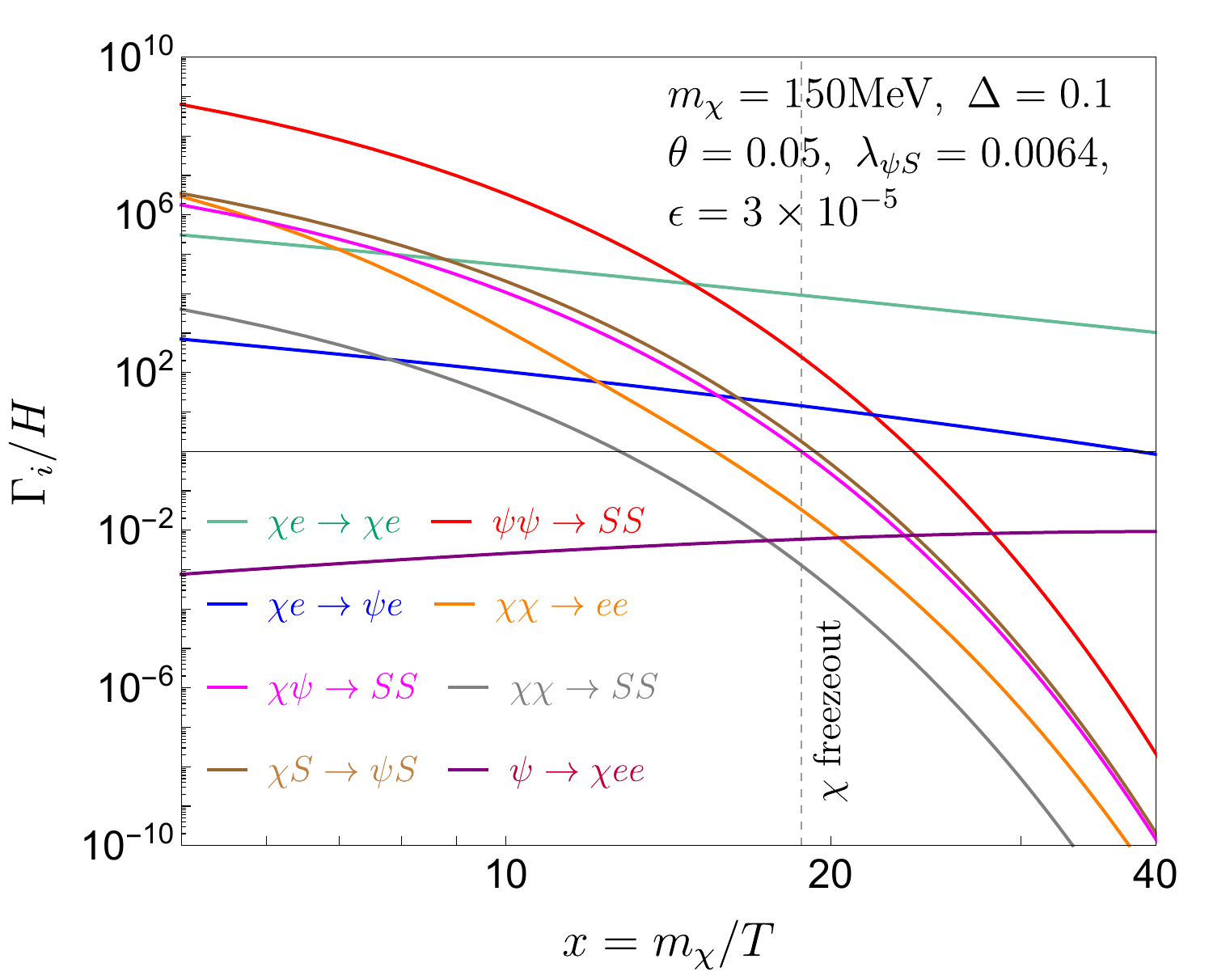}
\end{center}
\vspace{-.3cm}
\caption{\small Thermal rates of all relevant interactions. We plot the momentum transfer rate for elastic scattering (green) and $(n_\psi^{\rm eq})^2 \left< \sigma v \right>/n_\chi^{\rm eq}$ for $\psi$ annihilations (red). Freeze-out is defined by $n_\chi/n_\chi^{\rm eq}=2.5$. The parameters are given by $m_\chi=150\text{ MeV},~\Delta=0.1,~\theta=0.05, ~\epsilon=3\times10^{-5},~\lambda_{\psi S}=0.0064$, $\alpha_d=0.5$, $m_S=156\text{ MeV}$, and $m_A=3m_\chi$. 
\label{fig:rates2}}
\end{figure}

The main reactions involving the DM particle $\chi$ are
\begin{itemize}
\item visible and dark annihilations: $\chi\chi^*\to e^+e^-$, $\chi\chi^*\to SS$, $\chi\psi\to SS$
\item visible and dark exchange reactions: $\chi e \to \psi e$, $\chi S\to \psi S$, $\chi e^+ e^- \rightarrow \psi$
\item visible and dark elastic scatterings: $\chi e\to \chi e$, $\chi S\to \chi S$
\end{itemize}
Furthermore the final $\chi$ abundance also depends on the annihilation rate of the dark partner $\psi$: $\psi\psi^*\to SS$ and  $\psi\psi^*\to e^+e^-$.  The rates of most of these reactions are displayed in Fig.~\ref{fig:rates2} for an example parameter point.

To derive simple estimates, we work in the limit of small $\theta$ and $m_e / m_\chi$, and large $m_A / m_\chi$.  We neglect $O(1)$ factors arising from ratios of $\psi, \chi$, and $S$ masses.  The thermal averages of the above processes can be written as:
\begin{align}\label{estimates}
&\langle\sigma_{\chi\chi\to  e^+e^-} v \rangle\simeq \frac{e^2g_D^2\epsilon^2}{\pi}\frac{m_\chi^2}{m_A^4} \frac{T}{m_\chi}\equiv \frac{T}{m_\chi}\sigma_\chi \, , \\\label{estimates2}
&\langle\sigma_{\psi e\to  \chi e} v\rangle\simeq \frac{e^2g_D^2\epsilon^2\theta^2}{8\pi}\frac{m_\chi^2}{m_A^4}\Delta^2\equiv \frac{\theta^2\Delta^2}{8}\sigma_\chi\, , \\
&\langle\sigma_{\psi S\to  \chi S} v\rangle, \langle\sigma_{  S S\to \psi \chi} v\rangle\simeq \frac{\lambda_{\psi S}^2\theta^2}{16\pi m_\psi^2}\equiv \theta^2\sigma_\psi\, ,
\end{align}
where we define
\be
\langle\sigma_{\psi \psi\to  S S}\rangle \simeq\frac{\lambda_{\psi S}^2}{16\pi m_\psi^2}\equiv \sigma_\psi
\ee
Notice that $\chi$ annihilations to electrons are $p$-wave suppressed and we assumed $\Delta \times x_f \gtrsim 1$ in the equation for $\chi e \to \psi e$. In the opposite limit, $\Delta\times x_f \lesssim 1$, the zero appearing in Eq.~\ref{estimates2} at $\Delta=0$ is lifted by the temperature dependence of the thermal average and 
\be\label{estimate3}
\langle\sigma_{\psi e\to  \chi e} v\rangle\simeq \frac{6 e^2g_D^2\epsilon^2\theta^2}{\pi}\frac{T^2}{m_A^4}\, .
\ee 
In the following we will keep using Eq.~\ref{estimates2} which is however only valid for $\Delta\times x_f \gtrsim 1$.

We have only included the process affecting the number density of $\chi$ and which are exothermic (in the parameter space discussed in the paper). Assuming thermal equilibrium the thermal average of the inverse endothermic processes are obtained by using detailed balance
\be
n_\psi^{\eq} \langle\sigma_{\psi X\to  \chi X} v\rangle = n_\chi^{\eq} \langle\sigma_{\chi X\to  \psi X} v\rangle
\ee
with $X=e,S$.

Coscattering occurs when various conditions are realized: the DM particle $\chi$ is in kinetic equilibrium with the SM (I), there is no chemical potential for $\chi$ (II), and the last reactions to decouple which changes the $\chi$ number density are exchange reactions $\chi\leftrightarrow \psi$ (III). While this is also the setup of Ref.~\cite{DAgnolo:2017dbv}, here we focus on the situation in which the dominant coscattering is $\chi e \to\psi e$, thanks to the enhancement coming from the large number density of relativistic electrons (IV).   All the above conditions have to be enforced at the freeze-out temperature $T_f$ when the exchange reactions in (III) leave equilibrium. 

The exchange of $\chi\leftrightarrow \psi$ from inverse decay is subdominant  to $\chi e \to\psi e$ when
\be\label{decay}
n_e \langle\sigma_{\psi e\to  \chi e} v\rangle \gtrsim  \Gamma_{\psi \rightarrow \chi e^+ e^-} \approx  \frac{e^2 g_D^2\epsilon^2 \theta^2\, (\Delta m)^5}{128\pi^3 m_A^4},
\ee
where we used detailed balance to rewrite the endothermic reactions in terms of exothermic ones.
Using Eq.~\ref{estimates2}, Eq.~\ref{decay} requires
\be \label{eq:NoInverse}
\frac{\Delta m}{m_\chi} \lesssim \frac{6}{x_f} \approx 0.4
\ee
For $\Delta=0.1$ ($\Delta m\approx 0.05$), as used for our numerical results, the contribution from inverse decay is always negligible.  
For much smaller values of $\Delta$, Eq.~\ref{estimates2} is not accurate and instead Eq.~\ref{estimate3} should be used, which leads to similar conclusions. Finally, we verified numerically that inverse decay is negligible for $\Delta \times x_f\sim 1$.

In order for (II) to be satisfied, it is sufficient for dark annihilation of $\psi$, $\psi\psi^*\to SS$, to be faster than the coscattering reaction, $\chi e\to \psi e$, at $T_f$:
\be
(n_\psi^{\eq})^2 \langle\sigma_{\psi \psi\to  S S}\rangle\gtrsim (n_\chi^{\eq}) (n_e^{\eq})  \langle\sigma_{\chi e\to  \psi e} v\rangle.
\ee
Using detailed balance and the expression for relativistic and non-relativistic number densities in kinetic equilibrium we obtain the following relation
\be\label{condII}
\frac{\sigma_\psi}{\sigma_\chi}\gtrsim e^{m_\psi/T_f}\theta^{2} \times c_{II}\, ,
\ee
with $c_{II}\approx 0.35\,\Delta^2/x_f^{3/2}$, expressing a lower bound on ${\sigma_\psi}/{\sigma_\chi}$.

In order for (III) to be satisfied, that is to be in the coscattering regime,  all annihilations for $\chi$ should be slower than $\chi e\to \psi e$ at $T_f$. Starting with $\chi\chi^*\to e^+e^-$, this requires
\be\label{condIIIone}
e^{(2m_\chi-m_\psi)/T_f}\theta ^2\times c_{III}^{(1)}\gtrsim 1\,,
\ee
with $c_{III}^{(1)}\approx 0.35 \,\Delta^2/x_f^{1/2}({m_\psi}/{m_\chi})^{3/2}$.
This relation can also be interpreted as the fact that coscattering allows the usual thermal annihilation cross section $\chi\chi^*\to e^+e^-$ to be exponentially smaller than its thermal freeze-out value, as it is apparent by rewriting Eq.~(\ref{omegaratio}) as $\sigma_\chi \approx \sigma_{\rm WIMP}\times \frac{1}{ B}$
where
\be\label{thermalboost}
B\equiv e^{(2m_\chi-m_\psi)/T_f}\theta ^2\times b\, \gtrsim 1\,.
\ee
with $b\approx 0.4\,\Delta^2 p^{-1}x_f^{-3/2}({m_\psi}/{m_\chi})^{3/2}$.

The two additional reactions $\chi\chi^*\to SS$ and $\chi\psi\to SS$ impose constraints that depend on the relation between $m_S$ and $m_{\chi, \psi}$. 
We discuss these constraints assuming that both reactions are endothermic, $m_S \ge (m_\chi+m_\psi)/2$. This is the relevant situation for the parameter choice made in the paper ($r_S=0.75$). From this choice it follows that the exponential suppression of the two thermal averages is the same, but $\chi\psi\to SS$ is only suppressed by one power of the mixing angle $\theta^2$. Requiring $\chi\psi\to SS$ to be out of equilibrium at $T_f$ is thus the leading constraint and it requires 
\be\label{condIIItwo}
\frac{\sigma_\psi}{\sigma_\chi}\lesssim e^{(2m_S-m_\psi)/T_f}\times c_{III}^{(2)}\, ,
\ee
with $c_{III}^{(2)}\approx1.44\,{\Delta^2}/{x_f^{3/2}}({m_\psi m_\chi}/{m_S^2})^{3/2}$.

Finally the requirement that coscattering occurs by scattering on electrons (IV) requires  $\chi S\to \psi S$ to be out of equilibrium at $T_f$:
\be\label{condIV}
\frac{\sigma_\psi}{\sigma_\chi}\lesssim e^{m_S/T_f}\times c_{IV},
\ee
where $c_{IV}\approx 0.72\,{\Delta^2}/{x_f^{3/2}}({m_\chi}/{m_S})^{3/2}$. Notice that if $m_S < m_\psi$, which we assumed, this constraint is subleading to Eq.~(\ref{condIIItwo}).

Putting all the conditions together, the ratio $\sigma_\psi/\sigma_\chi$ is constrained to lie in the interval
\be
c_{II}\,e^{{m_\psi}/{T_f}}\theta^{2}\lesssim \frac{\sigma_\psi}{\sigma_\chi}\lesssim c_{III}^{(2)}\,e^{(2m_S-m_\psi)/{T_f}}.
\ee
For fixed values of the masses, this can be interpreted as a constraint on $\theta$, $\theta^2 \lesssim c_{III}^{(2)}/c_{II} e^{-2(m_\psi-m_S)/T_f}$, and an upper bound on the boost factor
\be
B \lesssim b\, c_{III}^{(2)}/c_{II}e^{(2 m_\chi + 2 m_S - 3 m_\psi)/T_f}
\ee
Notice that for a fixed value of $r_S$ there is a maximal value of $\Delta m/m_\chi$ such that $B>1$.

\begin{figure*}[!!!t]
\begin{center}
\includegraphics[width=0.8 \textwidth]{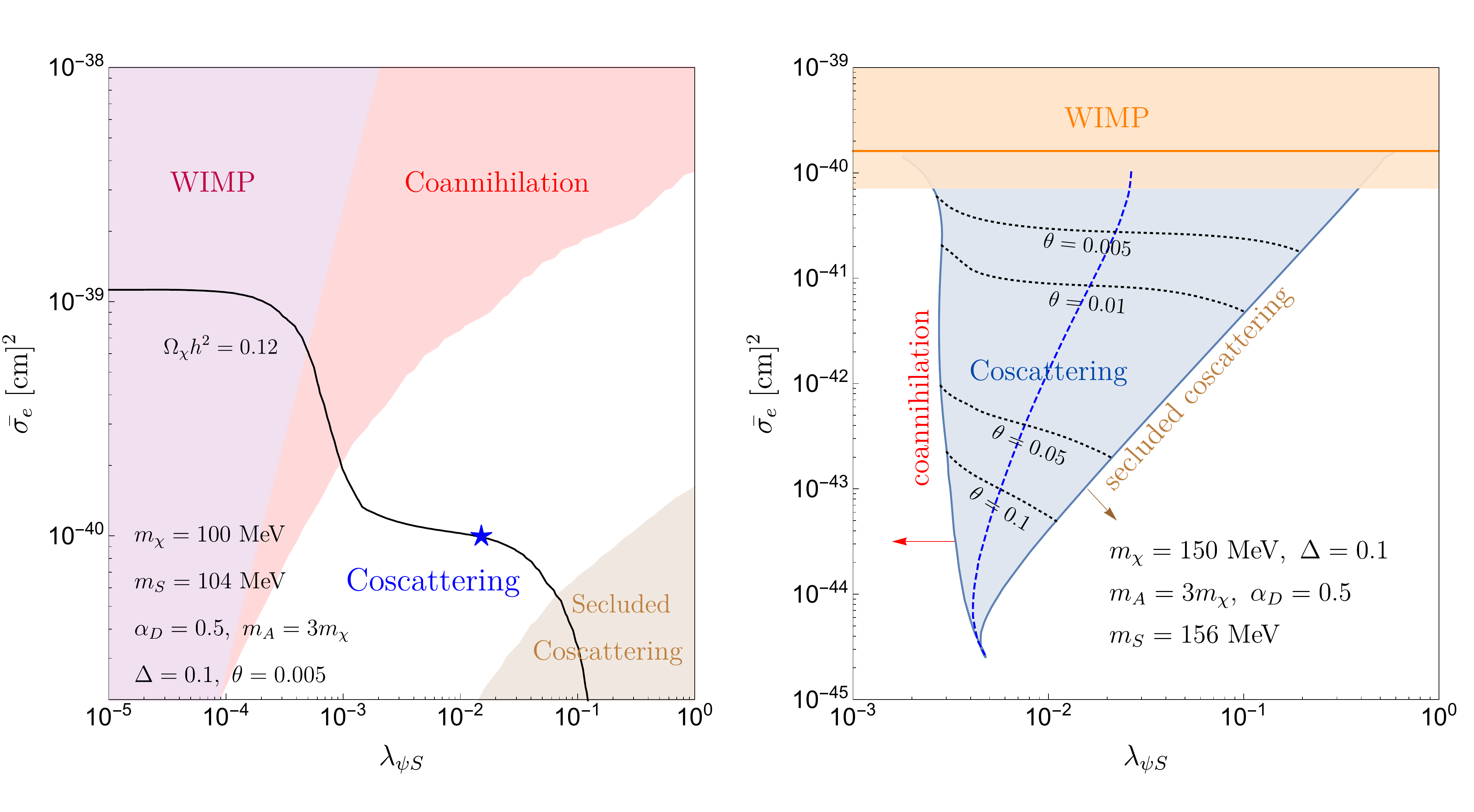}
\end{center}
\vspace{-.3cm}
\caption{\small {\emph{Left panel:}} freeze-out phase diagram for the scalar QED model. In each point of the plane a freeze-out temperature $T_f$ is defined to be the temperature at which $n_\chi/n_\chi^{\textrm{eq}}= 2.5$. A freeze-out phase is then assigned by comparing the rate of the various reactions involving $\chi$ as described in the text. On the black line the observed relic abundance is reproduced. The blue star corresponds to $\lambda^\star_{\psi S}=\sqrt{\lambda_{\min} \lambda_{\max}}$ where $\lambda_{\min}$ and $\lambda_{\max}$ correspond, for fixed $\theta$, to the minimal and maximal values of $\lambda_{\psi S}$ in the coscattering phase, respectively.  In the main body of the text we choose $\lambda_{\psi S} =\lambda^\star_{\psi S}$ in order to reside in the coscattering phase.  {\emph{Right panel:}}  for every point on the plane, $\theta$ is selected to reproduce the observed relic density. In the blue wedge, this occurs in the coscattering phase. The dashed blue line corresponds to the locus of points for which $\lambda_{\psi S} = \lambda^\star_{\psi S}$.
\label{fig:phases}}
\end{figure*}

{ \bf Freeze-out phases} Fig.~\ref{fig:phases} shows how the scalar QED model described in the paper may exhibit different DM freeze-out mechanisms depending on the choice of parameters. We display the phase diagram in the $(\lambda_{\psi S}, \bar\sigma_e)$ plane. At each point in the plane we solve the coupled Boltzmann equations for the number densities of $\chi$ and $\psi$, and we calculate the final abundance of $\chi$. A freeze-out temperature $T_f$ is defined as the SM temperature at which $n_\chi/n_\chi^{\rm{eq}}= 2.5$. At this temperature the rates of the various processes are compared to determine which mechanism determines the abundance. On the black line the final $\chi$ abundance corresponds to the observed DM density.

Notice that everywhere on the plane the ratio $\Gamma_{\chi e\to \psi e}/\Gamma_{\chi \chi \to e^+e^-}$ is fixed to be the left hand side of Eq.~(\ref{condIIIone}) and thus only depends on $T_f$. In particular the ratio is greater than one (meaning that $\chi e\to \psi e$ is faster than $\chi \chi \to e^+e^-$ at $T_f$) everywhere in the plane for the given choice of parameters. 

Starting from the left side of the plot, at small $\lambda_{\psi S}$, the first reaction to decouple is $\psi\psi^*\to SS$. This implies that the $\chi$ abundance is fixed as soon as $\chi^*\chi\to e^+ e^-$ annihilations decouple, which is the standard WIMP scenario. Notice that the black line representing the observed DM abundance is approximately horizontal in this region as a consequence of the fact that the final abundance only depends on the $\chi \chi^* \to e^+e^-$ cross section.

Moving to the right, for fixed $\bar \sigma_e$, we hit the boundary after which $\psi\psi^*\to SS$ decouples after $\chi \chi^*\to e^+e^-$ but before ${\chi e\to \psi e}$. This is the classical coannihilation scenario~\cite{DAgnolo:2018wcn, Griest:1990kh} in which the abundance of the inert state $\chi$ is depleted by the existence of another state $\psi$ with larger interactions and with which $\chi$ shares a chemical potential (this condition being enforced by fast $\chi e \to \psi e$ exchange reactions).  Notice that the black line is approximately vertical in this region as a consequence of the fact that freeze-out is determined by the decoupling of $\psi\psi^*\to SS$.

Increasing $\lambda_{\psi S}$ even more, $\psi\psi^*\to SS$ becomes faster than both $\chi\chi^*\to e^+e^-$ and $\chi e \to \psi e$, which leads to the coscattering regime. In the white part of the left of Fig.~\ref{fig:phases}, the final abundance is determined by inelastic scattering on electrons through $\chi e \to \psi e$, implying that iso-countours of $\chi$ abundance are independent of $\lambda_{\psi S}$. Finally moving to even larger $\lambda_{\psi S}$, in the brown shaded region, the reaction $\chi S\to \psi S$ becomes the dominant one and sets the final abundance; this is the regime of secluded coscattering discussed in Ref.~\cite{DAgnolo:2017dbv}.

Notice that moving from left to right, points on the black line (an iso-contour of $\chi$ relic abundance) have monotonically decreasing direct detection cross section.

Notice also that moving to larger values of $\bar \sigma_e$, the boundaries between the various phases all move to larger values of $\lambda_{\psi S}$, as a consequence of the fact that for larger $\bar \sigma_e$, both $\Gamma_{\chi e\to \psi e}$ and $\Gamma_{\chi \chi \to e^+e^-}$ are larger, shifting the transition values of $\lambda_{\psi S}$ to larger coupling sizes.  The blue star shows the value of $\lambda_{\psi S}$ chosen in the main body of the text in order to reside within the coscattering phase.

In the right panel of Fig.~\ref{fig:phases} we show the phases of the model in the same plane, but varying $\theta$ point by point to fix the right relic density. As $\theta$ increases coscattering and annihilations of $\chi$ and $\psi$ with the hidden sector state $S$ grow, reducing the available parameter space for coscattering off electrons. At the same time the coupling to electrons required to reproduce the measured relic density decreases. When this coupling becomes large we exit the coscattering regime to enter a WIMP phase where $\chi\chi^* \to e^+e^-$ sets the relic density.

{ \bf Direct detection}
The direct detection scattering cross section of $\chi$ on electrons, $\bar \sigma_e$, is given by~\cite{Berlin:2018bsc}
\be\label{dde}
\bar\sigma_e \simeq  \frac{e^2g_D^2\epsilon^2 c_\theta^4}{\pi} \frac{m_e^2}{m_A^4}= \frac{m_e^2}{m_\chi^2}\sigma_\chi.
\ee
where we assumed $m_\chi\gtrsim m_e$.

Eqs.~(\ref{dde}) and~(\ref{estimates}) define a one-to-one relation, $\bar\sigma_e^{\rm WIMP}(m_\chi)$, between the mass of a WIMP annihilating to electrons and the size of its direct detection cross section.
For a coscattering DM candidate, $\bar\sigma_e^{\rm WIMP}(m_\chi)$ is a strict upper bound on the size of the direct detection cross section and, in particular, $\bar\sigma_e(m_\chi)/\bar\sigma_e^{\rm WIMP}\approx 1/B$.

For a given mass, $m_\chi$, the upper bound on $B$, described in the previous section, corresponds to the smallest possible value of the direct detection cross section (see Fig.~\ref{fig:pheno}).

\bibliographystyle{apsrev4-1}
\bibliography{biblio}

\end{document}